\documentclass[%
 reprint,
nofootinbib,
 amsmath,amssymb,
 showkeys,
 aps,
]{revtex4-1}

\usepackage{graphicx}% Include figure files
\usepackage{dcolumn}% Align table columns on decimal point
\usepackage{bm}% bold math
\usepackage{hyperref}% add hypertext capabilities
\usepackage{enumerate}% add hypertext capabilities
\usepackage[utf8]{inputenc}

\def\i{{\rm i}}

\def\GMc2{{\rm G M_{\odot} c^{-2}}}

\usepackage{color}
\definecolor{purple}{rgb}{0.8,0.4,0.8}

\usepackage{color}

\begin{document}

\title{Axion star collisions with black holes and neutron stars in full 3D numerical relativity}

\author{Katy Clough$^1$}
 \email{katy.clough@phys.uni-goettingen.de}
\author{Tim Dietrich$^2$}
 \email{diettim@nikhef.nl}
  \author{Jens C. Niemeyer$^1$}
 \email{jens.niemeyer@phys.uni-goettingen.de}
\affiliation{$^1$Institut f{\"u}r Astrophysik, Georg-August 
  Universit{\"a}t, Friedrich-Hund-Platz 1, D-37077 G{\"o}ttingen, Germany}
  \affiliation{$^2$ Nikhef, Science Park, 1098 XG Amsterdam, The Netherlands} 

\begin{abstract}
Axions are a potential dark matter candidate, which may condense and 
form self gravitating compact objects, called axion stars (ASs).
In this work, we study for the first time head-on collisions of relativistic 
ASs with black holes (BHs) and neutron stars (NSs).
In the case of BH-AS mergers we find that, in general, 
the largest scalar clouds are produced by mergers of low compactness ASs and spinning BHs. 
Although in most of the cases which we study the majority of 
the mass is absorbed by the BH within a short time after the merger, 
in favourable cases the remaining cloud surrounding the final BH remnant 
can be as large as $30\%$ of the initial axion star mass, 
with a bosonic cloud mass of $\mathcal{O}(10^{-1})M_{\rm BH}$ and peak energy density 
comparable to that obtained in a superradiant build up.
This provides a dynamical mechanism for the formation of long lived scalar hair, 
which could lead to observable signals in cases where the 
axion interacts with baryonic matter around the BH, 
or where it forms the seed of a future superradiant build up in highly spinning cases.
Considering NS-AS collisions we find two possible final states 
(i) a BH surrounded by a (small) scalar cloud, 
or (ii) a stable NS enveloped in an axion cloud of roughly the same mass as the initial AS. 
Whilst for low mass ASs the NS is only mildly perturbed by the collision, 
a larger mass AS gives rise to a 
massive ejection of baryonic mass from the system, purely due to gravitational effects. 
Therefore, even in the absence of a direct axion coupling to baryonic matter, 
NS-AS collisions could give rise to electromagnetic observables 
in addition to their gravitational wave signatures.
\end{abstract}

\maketitle

\section{\label{sec:intro}Introduction}

In the wake of multiple LIGO detections~\cite{Abbott:2016blz,Abbott:2016nmj,Abbott:2017vtc,Abbott:2017oio,Abbott:2017gyy},
including GW170817~\cite{TheLIGOScientific:2017qsa}, the first
combined detection of GWs and electromagnetic signals from 
the same astrophysical source,
there has been renewed interest in the simulation 
of exotic compact objects (ECOs) which could mimick BH or NS observations, or provide altogether new,
and as yet undetected, observational signatures \cite{Giudice:2016zpa}.

One of the simplest potential ECOs is a boson star (BS), which is a stable solitonic solution to the coupled Einstein-Klein-Gordon 
equations for a complex scalar field with gravity. 
The idea of a self gravitating field configuration dates back to proposals by Wheeler for ``geons"~\cite{Wheeler:1955zz}, but was first shown to work for complex scalar fields in~\cite{Kaup:1968zz, Ruffini:1969qy}. 
The ideas were extended to real massive scalar fields in~\cite{Seidel:1991zh}, with the (quasi) stable objects later dubbed oscillotons. 
A non trivial self interaction potential, motivated by, for example, low energy effective theories from string theory or other models, 
modifies the stability and profile of the solutions, giving rise to new classes of self interacting BSs~\cite{Eby:2015hsq, Colpi:1986ye, Krippendorf:2018tei}. 

The standard model of particle physics does not contain a bosonic particle that would allow the formation of BSs. Dark matter, on the other hand, may well be composed of bosons, with axion-like particles (ALPs) being a well-motivated class of candidates (see~\cite{Marsh:2015xka} for a thorough review). ALPs are very light, weakly coupled particles that are produced with practically vanishing momenta and extremely high occupation numbers. They are usually treated as classical, real scalar fields, subject to a cosine potential parametrised by the axion decay constant $f_a$ and the axion mass $m_a$. The leading order $\phi^4$ self-interaction for axions is thus attractive, but whilst here we simulate the full cosine potential, we use a large value of $f_a$ such that the the axion is effectively a massive boson and self interactions are negligible. In future work we hope to expand the study to quantify the effect of increasing self interactions.

In this work, we focus on ASs with masses comparable to the objects with 
which they collide since such setups will give rise to strong GW signals.
Consequently, the AS masses in the NSAS configurations are $M \sim M_\odot$, 
whilst in the BHAS case the physical mass of the BH sets the scale of the simulation such that the interpretation as solar mass BHs corresponds to solar mass ASs, and for supermassive BHs similarly the ASs have masses up to $M \sim 10^{10} M_\odot$. With the AS mass fixed, we choose values for the axion mass $m_a$ such that the star is comparatively compact - ie, such that the de Broglie wavelength of the axion is comparable to the radius of the BH or NS. In the solar mass case this corresponds to $m_a \sim 10^{-10}$eV (the lower end of the QCD axion scale), and for supermassive BHs, as low as $m_a \sim 10^{-20}$eV (corresponding to ultra-light ALPs).

The original QCD axion emerges as a consequence of the Peccei-Quinn (PQ) symmetry breaking mechanism to solve the strong CP problem \cite{Peccei:1977hh, Weinberg:1977ma}. Its mass is constrained to lie between $m_a = 10^{-12}$ eV and $10^{-2}$ eV, with decay constant $f_a \ll M_{pl}$. If dark matter consists of QCD axions whose PQ symmetry was broken after inflation, high-amplitude density fluctuations on the scale of the cosmological horizon at the time of the QCD phase transition are predicted to exist. They collapse during the radiation era and form so-called axion miniclusters with typical masses of $\sim 10^{-11} M_\odot$ \cite{Hogan:1988mp,Kolb:1993zz}. A fraction of these masses may have formed ASs either directly during the collapse \cite{Schive:2014dra,Veltmaat:2018dfz} or by scalar wave condensation \cite{Levkov:2018kau}. The subsequent evolution of the AS mass function as a result of minicluster mergers or ongoing condensation is largely unknown; although their typical masses at formation are well below those we consider in this work, the existence of a high-mass tail of the distribution cannot be ruled out at present. A population of relativistic ASs could also have been produced by non-standard primordial perturbations with enhanced small-scale power \cite{Widdicombe:2018oeo}. 

Ultra-light ALPs with masses in the range of $\sim 10^{-22} - 10^{-20}$ eV, motivated from string theory compactifications \cite{Svrcek:2006yi, Arvanitaki:2009fg}, are candidates for ``fuzzy dark matter'' (FDM) with interesting new phenomenology on scales of their de Broglie wavelength in dark matter halos on scales of kpc \cite{Hu:2000ke, Hui:2016ltb}. In particular, cosmological simulations using the nonrelativistic Schrödinger-Poisson equations to describe FDM show the formation of ASs in the form of solitonic halo cores \cite{Schive:2014dra,Veltmaat:2018dfz} with masses of the order of $10^{7} M_\odot$ and above. The evolution of their mass function as a result of halo mergers was studied in \cite{Du:2016aik} and shown to approach the core-halo mass relation found by \cite{Schive:2014dra}. 

Thus, even if axions make up all the dark matter, relativistic ASs would constitute only a small ($< 1 \%$) fraction of the total mass (also, note that observational constraints from microlensing on compact ASs would be similar to those on primordial BHs). Nevertheless, there could still be a sufficient number density for collisions with known objects such as BHs and NSs to occur, motivating an exploration of their observable signatures.

Simulations of ECO mergers more generally have to date focused mainly 
on complex scalar field BSs, 
e.g.~\cite{Balakrishna:1999sv,Palenzuela:2006wp, Choptuik:2009ww, Bezares:2017mzk, Lai:2004fw, Palenzuela:2007dm, Cardoso:2016oxy,Dietrich:2018bvi}, 
but other classes such as real scalar field oscillotons \cite{Helfer:2018vtq}, and just recently, 
massive, complex vector field Proca stars in \cite{Sanchis-Gual:2018oui,Sanchis-Gual:2017bhw} have been explored. 

There has, to the best of our knowledge, not yet been a study of the merger of ASs with BHs and NSs, even though 
these collisions are of interest since
the ASs could act as a potential NS or BH mimickers in GW signals. In addition, 
since some types of axions are expected to couple weakly to 
baryonic matter, such mergers could result in distinct 
multi-messenger events, with, for example, phenomena such as Fast Radio Bursts (FRBs).
In the case of a NS-AS collision the axions may interact with 
the neutron star matter either during or after the merger (see for example 
\cite{Hook:2017psm, Eby:2017xaw, Raby:2016deh, Iwazaki:2014wka, Barranco:2012ur}
and the discussion in~\cite{Dietrich:inprep}). 
Similarly, the case of a BH-AS merger 
may provide a mechanism by which one can dynamically form long lived (quasibound) scalar clouds, or ``wigs'', around BHs 
\cite{Barranco:2012qs, Barranco:2017aes, Sanchis-Gual:2016jst, Herdeiro:2015waa, Hod:2012px}. 
Such clouds could then interact with any baryonic matter present in a potential accretion disc, or, in appropriate cases,
provide the seeds for a superradiant instability (see e.g. \cite{Brito:2015oca, Brito:2014wla}) to develop. 
Several other novel methods for detecting such scalar clouds have also been proposed, 
see e.g. \cite{Ferreira:2017pth, Hannuksela:2018izj, Degollado:2014vsa}.
Consequently, these events have great potential to further 
constrain the properties of the axion sector.

For our simulations, we assume that the axion field is only 
coupled to baryonic matter via the gravitational interaction, i.e., 
they interact due to their mutual impact on the metric, and no 
additional couplings are implemented.
As discussed, we also focus on relatively high values for the axion decay constant of $f_a =0.5 M_{pl}$, meaning that self interactions are negligible.
These choices are most relevant to the case of ALPs with a 
negligible coupling to standard model physics. 

The paper is organised as follows.
Sec.~\ref{sec:ICs} summarises the numerical methods used for
the work and presents code tests, as well as a preliminary investigation 
to determine the initial conditions for the simulations shown in the 
remainder of the paper.
In Sec.~\ref{sec:resultsBH} we consider the impact of
varying the AS compactness on the remnant clouds left by BH-AS 
from mergers of ASs with spinning and non-spinning BHs. 

In Sec.~\ref{sec:resultsNS} we discuss NS-AS collisions, and 
investigate the remnant of the head-on collision as the NS-AS mass ratio is varied. 
We summarize our results and future plans  in Sec.~\ref{sec:conclusions}.\\

Throughout this paper we employ geometric units, with $G=c=1$ and a mass scale $M$ 
which in the case of NSs is set to $M_\odot$, but in the case of BHs
is a free mass scale by which the results may be scaled for varying BH masses. 
Consequently code units for lengths, times and masses are multiples of the mass scale $M$\footnote{Note that the scale $\mu$ which appears in the potential function is the quantity
$\mu = m_a c / \hbar$, with dimension $[L^{-1}]$ such that a value of $\mu=1$ in code units corresponds to 
a particle mass of $m =1.3\times 10^{-10} ~ {\rm eV}$ in the NS case (note that $\hbar \neq 1$ in geometric units), 
we refer to~\cite{Dietrich:2018bvi} for further discussion.}.

\section{\label{sec:ICs}{Numerical methods}}

In this section we describe the set ups used for the respective simulations. We use the GRChombo code \cite{Clough:2015sqa} for the BH-AS evolutions, 
and the BAM code \cite{Brugmann:2008zz,Thierfelder:2011yi,Dietrich:2018bvi} for the NS-AS cases, but undertake comparisons to ensure they give consistent results.

\subsection{BH-AS collisions using GRChombo}

The GRChombo code~\cite{Clough:2015sqa} (\url{www.grchombo.org})
is based on the method-of-lines with fourth order spatial finite differences and fourth order explicit Runge-Kutta timestepping. 
It is built on top of the Chombo \cite{Chombo} framework for solving partial differential equations 
with adaptive mesh refinement (AMR). 
It supports non-trivial “many-boxes-in-many-boxes”
mesh hierarchies using block structured Berger Rigoutsos
grid generation \cite{BergerRigoutsis91}.
The grid is made out of a hierarchy of
cell-centered Cartesian grids consisting of a maximum of $L$ refinement levels
labeled $l = 0,...,L-1$ with resolution increasing by a factor of $2$ on successive levels according to $h_l= 2^{-l} h_0$. 
Given that the mesh is adaptive, the configuration of grids changes dynamically during the simulation. Around the BH
all refinement levels are utilised, but for the scalar matter the number required varies between 4 and 9. 
The regridding is determined by setting the thresholds $t_\phi$, $t_\chi$, and cut offs $\epsilon_\phi$, $\epsilon_\chi$ in the regridding condition:
\begin{multline}
\max 
\left( \frac{\Delta x}{t_\phi} \sqrt{ \sum_{ij} 
 \frac{ (\partial_i \partial_j \phi)^2 }{\epsilon_\phi +  |\partial_i \phi \partial_j \phi|} +
 \frac{ (\partial_i \partial_j \Pi)^2 }{\epsilon_\phi + |\partial_i \Pi \partial_j \Pi|} } ~ , 
 \right. \\
 \left. 
 \frac{\Delta x}{t_\chi} \sqrt{ \sum_{ij}
 \frac{ (\partial_i \partial_j \chi)^2 }{\epsilon_\chi + |\partial_i \chi \partial_j \chi|}}
\right) > 0.5
\end{multline}
such that smaller values of the thresholds force regridding for 
less varying/oscillatory data. In principle this allows us to maintain a
consistent level of refinement on the scalar matter around the BH, even as it decreases in amplitude by several orders of magnitude 
during the merger; see \cite{Chombo, Clough:2015sqa} for further details on the regridding process. 
The Berger-Oliger scheme is employed~\cite{Berger:1984zza} to coordinate time stepping on the grid hierarchy 
and we use a Courant-Friedrichs-Lewy factor of $0.2$ on each level.
The evolution of the scalar fields is based on the 
Klein-Gordon Equation and uses the methods outlined 
in Ref.~\cite{Helfer:2016ljl}.
The exact grid configurations for the three resolutions are given explicitly in Tab.~\ref{tab:GRChombo_grid}.

\begin{table}[t]
  \centering    
  \caption{Grid configurations for the GRChombo simulations for spinning BHs. 
    The columns refer to: $L$ total number of levels, 
    $n$ number of points along each dimension on the coarsest level, 
    $t_{\phi}$ the regridding threshold on $\phi$ (for cases $\phi_0 = 1/\sqrt{8 \pi} \ (0.20,~0.07,~0.02$), 
    $t_{\chi}$ the regridding threshold on $\chi$, 
    $\epsilon_{\phi}$ the regridding cutoff on $\phi$, 
    $\epsilon_{\chi}$ the regridding cutoff on $\chi$, 
    $h_0$ coarsest grid spacing, and
    $h_L$ finest grid spacing.} 
  \begin{tabular}{l|ccccccccc}        
    \hline
    Name & $L$ &  $n$ & $t_{\phi}$ & $t_{\chi}$ & $\epsilon_{\phi}$ & $\epsilon_{\chi}$ & $h_0$ & $h_L$ \\
    \hline
    R1    & 8   & 128 & (1.0,0.20,0.06) & 0.48 & 3e-6 & 1e-2  & 8.00  & 0.031250  \\
    R2    & 8   & 192 & (0.6,0.16,0.04) & 0.32 & 3e-6 & 1e-2  & 5.33  & 0.020833  \\
    R3    & 8   & 256 & (0.5,0.10,0.03) & 0.24 & 3e-6 & 1e-2  & 4.00  & 0.015625  \\    
    \hline
  \end{tabular}
  \label{tab:GRChombo_grid}
\end{table}

\subsection{NS-AS collisions using BAM}

The BAM code~\cite{Brugmann:2008zz,Thierfelder:2011yi,Dietrich:2015iva,Bernuzzi:2016pie,Dietrich:2018bvi} 
is based on the method-of-lines, Cartesian grids and finite
differencing. The grid is made out of a hierarchy of
cell-centered nested Cartesian boxes consisting of $L$ refinement levels
labeled $l = 0,...,L-1$ with increasing resolution. 
Each level's resolution increases by a factor of two
leading to a resolution of $h_l= 2^{-l} h_0$. 
The inner levels employ $n_{\rm mv}$ 
points per direction and move following the technique of `moving-boxes' 
where the star's center are estimated by the minimum 
of the conformal factor $\chi$. 
The outer levels remain fixed and employ 
$n^3$ grid points. Because of the symmetry of the problem, 
we employ bitant-symmetry to reduce the number of grid points and 
the computational costs by a factor of two. 
For the time integration a fourth order explicit Runge-Kutta type integrator 
is used with a Courant-Friedrichs-Lewy factor of $0.2$.
For the time stepping of the refinement level 
the Berger-Oliger scheme is employed~\cite{Berger:1984zza}. 
Metric spatial as well as scalar field derivatives are 
approximated by fourth order finite differences. 
The general relativistic hydrodynamic equations are solved with standard
high-resolution-shock-capturing schemes based on primitive reconstruction
and the local Lax-Friedrich central scheme for the numerical
fluxes; see~\cite{Thierfelder:2011yi,Bernuzzi:2012ci} for more details.
The evolution of the scalar fields based on the 
Klein-Gordon Equation uses techniques outlined 
in Ref.~\cite{Dietrich:2018bvi}.
The exact grid configurations are given 
explicitly in Tab.~\ref{tab:BAM_grid}.

For the construction of the initial NS-AS configurations 
we use BAM's multigrid solver. The solver employed the Conformal 
Thin Sandwich (CTS) formalism to obtain initial configurations 
in agreement with Einstein's field equations. 
However, we note that as outlined in~\cite{Dietrich:2018bvi} 
we do not resolve the equations for general 
relativistic hydrodynamics or the Klein-Gordon equation 
which would be required to obtain fully consistent 
initial configurations. 

\begin{table}[t]
  \centering    
  \caption{Grid configurations for the BAM simulations. 
    The columns refer to: $L$ total number of levels, 
    $n$ number of points along each dimension, 
    $L_{\rm mv}$ number of moving box levels using $n_{\rm mv}$ points per direction, 
    $h_0$ coarsest grid spacing, and
    $h_L$ finest grid spacing.} 
  \begin{tabular}{l|ccccccc}        
    \hline
    Name & $L$ &  $n$ & $L_{\rm mv} $ & $n_{\rm mv}$ & $h_0$ & $h_L$ \\
    \hline
    R1    & 7   & 160 & 4 & 96  & 16.00  & 0.25  \\
    R2    & 7   & 240 & 4 & 144 & 10.67  & 0.167 \\
    R3    & 7   & 320 & 4 & 192 & 8.00   & 0.125 \\    
    \hline
  \end{tabular}
  \label{tab:BAM_grid}
\end{table}

\subsection{Code testing and comparison}
\label{sec:testing}

\begin{figure}[t]
    \centering
    \includegraphics[width=0.49\textwidth]{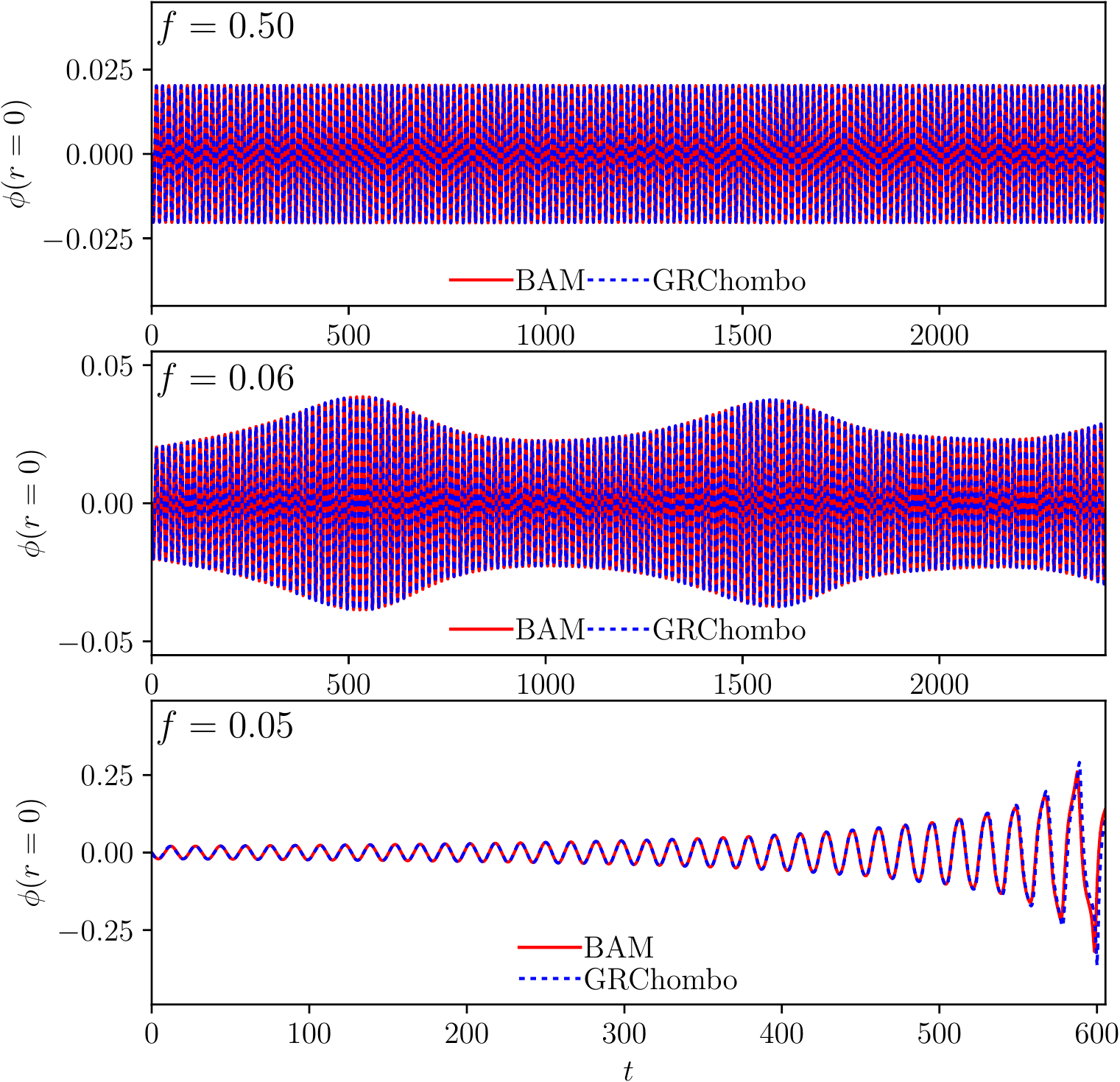}
    \caption{Single axion star comparison between GRChombo (blue) and BAM (red).
    We employ three different axion star potentials varying $f_a=0.50,0.06,0.05$.}
\label{fig:comparison_AS}
\end{figure}

For all presented configurations we performed
numerical simulations with different resolutions to 
provide an error estimate. In addition, we have validated our results by 
comparing directly the outcome of the GRChombo and BAM codes. 
For this purpose we evolve a set of single star AS configurations 
as well as one example BH-AS merger. We also refer to
Ref.~\cite{Dietrich:2018bvi} for a comparison of a BS-BS head-on 
merger between BAM and GRChombo. \\

Starting with isolated AS simulations, in Fig.~\ref{fig:comparison_AS}
we present the time evolution of the central value of the scalar field 
for a number of axion star potentials of the form
\begin{equation}
V(\phi) =  f_a^2 \mu^2 \left(1 - \cos\left(\frac{\phi}{f_a}\right)\right). \label{eq:V_axion}
\end{equation}
Here $f_a$ is the axion decay constant and $\mu = m_a c / \hbar$ is an inverse length scale which is a function of the 
axion mass $m_a$. 

For the comparison we picked an initial configuration with 
$\phi_c(t=0)=0.0,\pi_c(t=0)\approx-0.021$. 
For an axion decay constant of $f_a=0.50$ 
we find sinusoidal oscillations between 
$\phi_c = [-0.021,0.021]$ (top panel). 
Once we decrease the axion decay constant to $f_a=0.06$
the AS undergoes additional modulations (middle panel). 
Finally for $f_a=0.05$ the pressure support is not strong enough to avoid 
gravitational collapse and a BH forms (bottom panel). 
For all three cases $\phi_c$ is almost identical for BAM and GRChombo
and we find excellent agreement. 

\begin{figure}[t]
    \centering
    \includegraphics[width=0.49\textwidth]{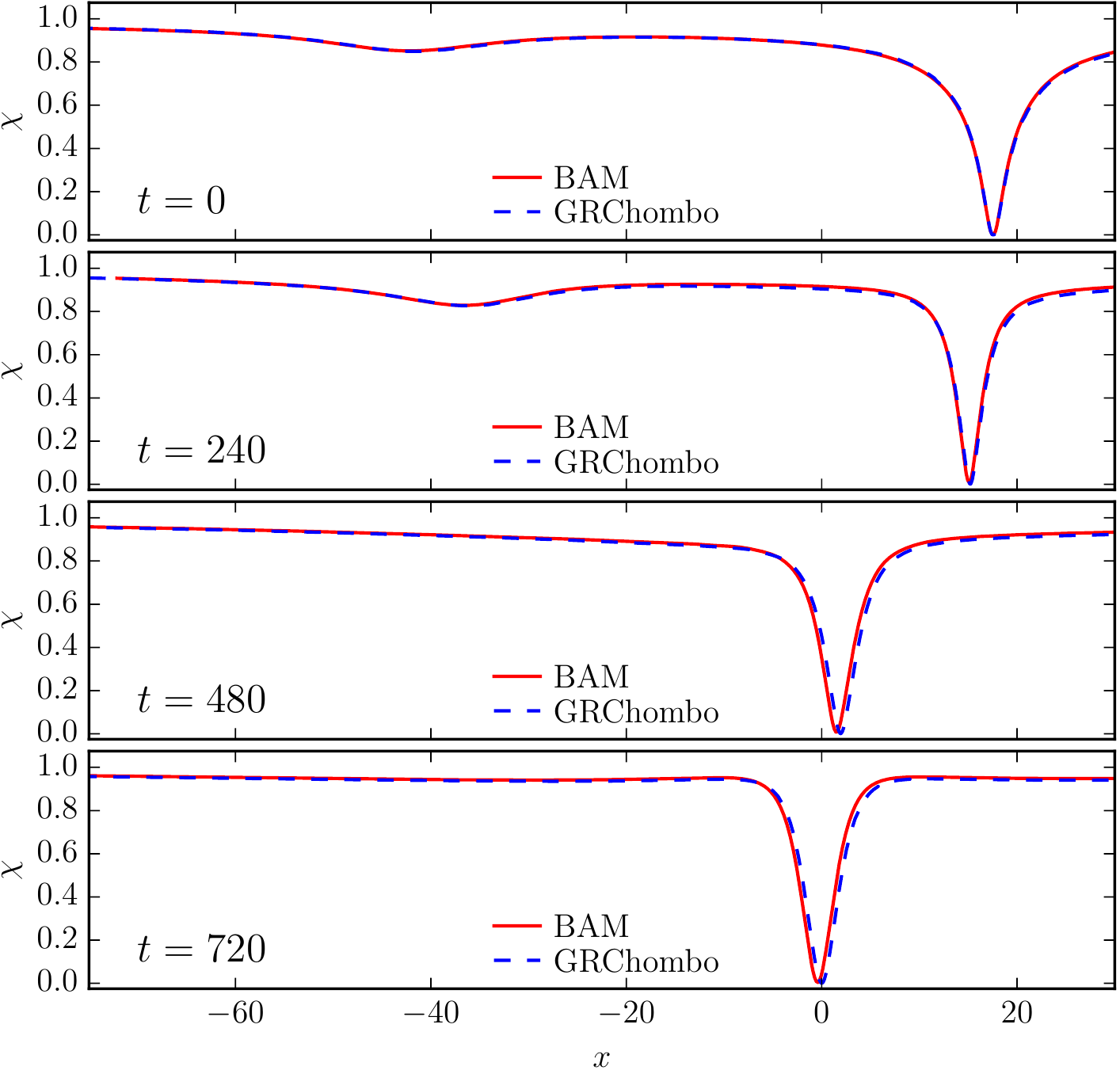}
    \caption{Spatial profile of the conformal factor $\chi$ for BAM (blue)
    and GRChombo (red) for different instances of time for a BH-AS collision. }
\label{fig:comparison_BHAS}
\end{figure}

More challenging than the single star comparison is the evolution of a 
BH-AS merger simulation. For this purpose we perform a head-on collision 
with an initial separation of $d=60$, an axion potential with $f_a=0.5$, 
and an AS with an initial central field value of $\phi_c(t=0)=0.0,\pi_c(t=0)=-0.021$ 
(as for the single AS simulations presented above). 
For comparison we present, although gauge dependent, 
the spatial profile of the conformal factor $\chi(x)$ 
for different times (Fig.~\ref{fig:comparison_BHAS}). 
We find overall good agreement between BAM and GRChombo although 
details in the simulation scheme and methods are different as, e.g., 
the usage of the Z4c for BAM or the CCZ4 evolution scheme for GRChombo.

\subsection{Dependence on initial separation}
\label{sec:separation}

\begin{figure}[t]
    \centering
    \includegraphics[width=0.49\textwidth]{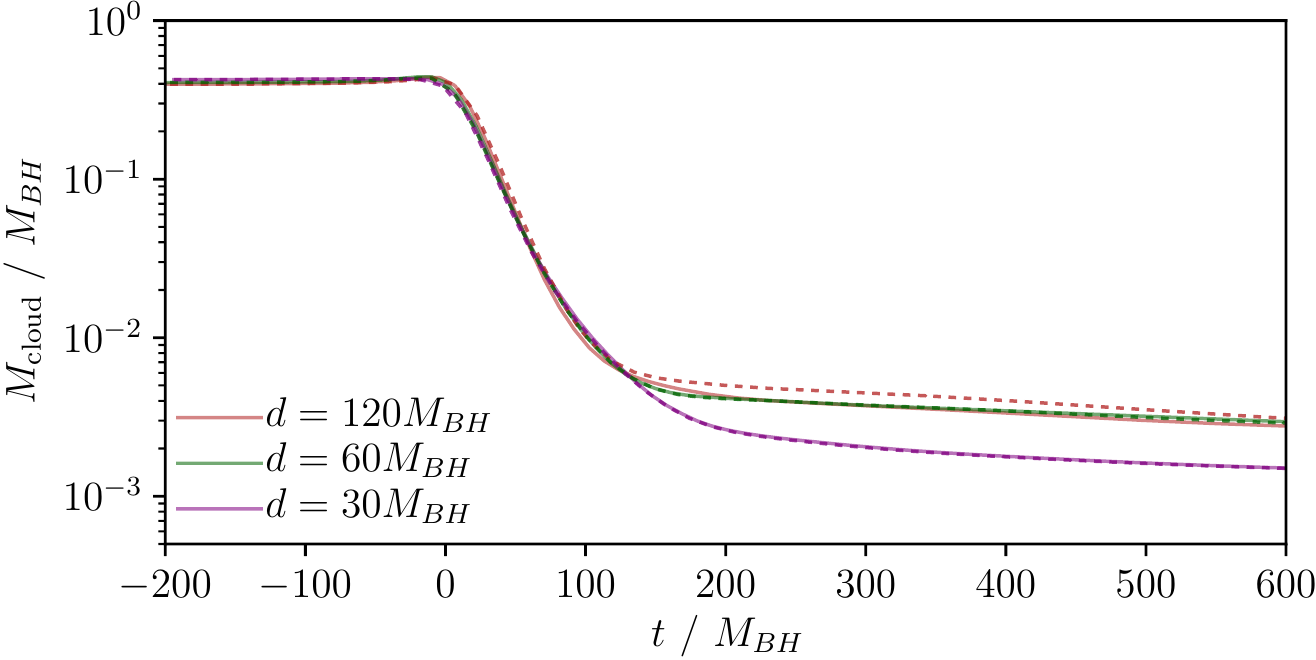}
    \caption{Evolution of the bosonic cloud mass surrounding the final BH remnant,
    for varying initial separations $d$, for an AS with $\phi_0=0.02$, 
    and a BH of mass
    $M_{\rm BH} = 1$ and $a=0$.
    We present results for resolution R2 with solid lines and corresponding results 
    for resolution R1 are shown dashed.
    Between $d=60M_{\rm BH}$ and $d=120M_{\rm BH}$ there is no significant difference in the cloud mass which forms.}
\label{fig:DistanceClouds}
\end{figure}

In order to determine for our simulations the optimal initial 
separation of the stars and BHs, we tested the impact of initial
separation on the bosonic cloud mass which formed in the 
case of a zero spin BH-AS merger.
The results are shown in Fig~\ref{fig:DistanceClouds}.

We find that the mass of the remnant cloud gets 
larger for an increasing initial separation. 
However, the absolute difference between large distances 
($d>60M_{\rm total}$) is small and below the uncertainty 
introduced by the numerical discretization, cf.~dashed and solid lines. 
We thus choose an initial separation of 100 $M_{\rm total}$, 
where $M_{\rm total}$ denotes the total mass of the system, 
i.e., the sum of the masses of the individual stars in isolation.

\section{\label{sec:resultsBH} BH-AS mergers}

In this section we study the head-on mergers of BHs and ASs. 
The BH can take any physical mass $M_{\rm BH}$, from a small primordial BH to supermassive BH, with the results scaled accordingly.
We take the mass of the AS to be of order the BH mass, and set $\mu=1$ in geometric units with $M_{\rm BH}=1$, such that the relevant axion mass varies according to $m_a \sim \left( \frac{M_{\rm BH}}{M_\odot} \right) 10^{-10}$ eV. This sets the radius of the axion star to be roughly of order the Schwarzschild radius of the BH, which is the regime which is most favourable for GW production. However, we mainly focus on ASs with a relatively low mass and compactness, since we expect (and indeed find) that these are the best candidates to form large scalar clouds around the final BH remnant.
Unlike in the case of superradiance, there is no physical requirement that $\mu ~ M \sim 1$, which essentially sets 
the de Broglie wavelength of the axion to the BH 
radius. In practise, it is numerically challenging to simulate 
ASs with vastly different sizes and masses to the BH, which is another reason that we focus in this work on such setups. 
These cases are of particular interest, since their remnants may go on to form a superradiant instability, but we emphasise that 
what we describe can be made more general, and does not require an exact mass correspondence or extremal spins.
Note that, as shown in \cite{Barranco:2012qs}, the dynamical resonances around the BH should be longer lived for cases in which $M \mu \ll 1$, in which case the half life of the scalar field can have a cosmological timescale. Given our choice of parameters, we should expect to see shorter lived solutions, but to confirm this we would need to evolve for longer periods beyond the merger, which is numerically challenging.

\subsection{Configurations}

For the study of BH-AS mergers, we perform simulations 
for 6 different binary configurations using three AS 
compactnesses and two BH spins, cf.~Table~\ref{tab:BHAS_setups}.

For the construction of the initial BH-AS configurations, we use the initial data for the ASs as in Ref.~\cite{Helfer:2016ljl}, 
with $\phi(x,y,z,t=0) = 0$ and $\Pi(x,y,z,t=0)$ set to an initial profile which is a stable solution for the free field oscilloton. 
Since we use a large value of $f_a$ (above the ``triple point'' as described in Ref.~\cite{Helfer:2016ljl}), the stars are stable
and far from collapsing to BHs, and the modulation of the base frequency due to the self interaction is small.

The AS's ADM (Arnowitt-Deser-Misner) mass is 
computed by fitting the Schwarzschild solution to the numerical solution
which is obtained via a shooting method, as in \cite{Helfer:2016ljl}. 
The bosonic mass of the AS is
determined by integrating the density of the field 
$E^{(\phi)}$ over the spatial volume of the initial slice.

We superpose this data with Bowen-York data \cite{BowenYork} for the spinning BHs, using the perturbative analytic solution 
for $\chi$ given in~\cite{Gleiser:1997ng}, and then apply a relaxation procedure (see \cite{Clough:2015sqa}) to
reduce the Hamiltonian constraint violation. Note that we first convert the AS data into 
conformally flat coordinates, so that it may be superimposed onto the conformally flat BH data with less deformation. 
The spin axis (the z-axis) is perpendicular to the merger direction (along the x-axis), and the initial distance is adjusted such that at $t=0$ it is 
approximately $d\approx100 M_{\rm total}$; 
cf. Sec.~\ref{sec:separation} for a study of the effect of 
different initial separations. 

\begin{table}[t]
  \centering    
  \caption{
    Simulated BHAS setups.
    The columns refer to: the name of the configuration, 
    the ADM mass of the BH in isolation $M_{\rm BH}$,  
    the AS's ADM mass in isolation $M_{\rm AS}$, 
    the AS's bosonic mass in isolation $M_{\rm AS}^*$,    
    the AS's compactness in isolation $\mathcal{C} = 2M_{\rm AS} / R$ 
    (where $R$ is the effective radius of the oscillaton which encompasses 95\% of its mass)
    the spin parameter for the BH $a = J/M$,
    the central value of the axion scalar field $\phi_c(t=0)$, 
    and the initial separation $d$ in units of $M_{\rm BH}$. } 
  \begin{tabular}{l|ccccccc}        
    \hline
    Name    & $M_{\rm BH}$ & $M_{\rm AS}$ & $M_{\rm AS}^*$ & $\mathcal{C}$ & $a$ & $\sqrt{8\pi}\phi_c^*$ & $d$ \\
    \hline
    ${\rm BHAS}_{\rm 0.02}^{0.0}$  & 1.00 & 0.20 & 0.20 & 0.014 & 0.0 & 0.02 & 120 \\
    ${\rm BHAS}_{\rm 0.07}^{0.0}$  & 1.00 & 0.36 & 0.35 & 0.048 & 0.0 & 0.07 & 136 \\
    ${\rm BHAS}_{\rm 0.20}^{0.0}$  & 1.00 & 0.52 & 0.52 & 0.130 & 0.0 & 0.20 & 152 \\  
    ${\rm BHAS}_{\rm 0.02}^{0.5}$  & 1.00 & 0.20 & 0.20 & 0.014 & 0.5 & 0.02 & 120 \\    
    ${\rm BHAS}_{\rm 0.07}^{0.5}$  & 1.00 & 0.36 & 0.35 & 0.048 & 0.5 & 0.07 & 136 \\  
    ${\rm BHAS}_{\rm 0.20}^{0.5}$  & 1.00 & 0.52 & 0.50 & 0.013 & 0.5 & 0.20 & 152 \\       
    \hline
  \end{tabular}
  \label{tab:BHAS_setups}
\end{table}

\subsection{Qualitative merger dynamics}

We find that the BH-AS merger provides a dynamical mechanism for the formation of scalar clouds around the BH.

We will see in the following section that we can classify the merger dynamics of NS-AS systems into three qualitatively different categories depending on the compactness (and mass) of the AS with which they collide. In the BH-AS cases there is no such distinction, but rather a gradual change in the formation of the bosonic clouds with AS compactness.

\begin{figure*}[t]
    \centering
    \includegraphics[width=0.99\textwidth]{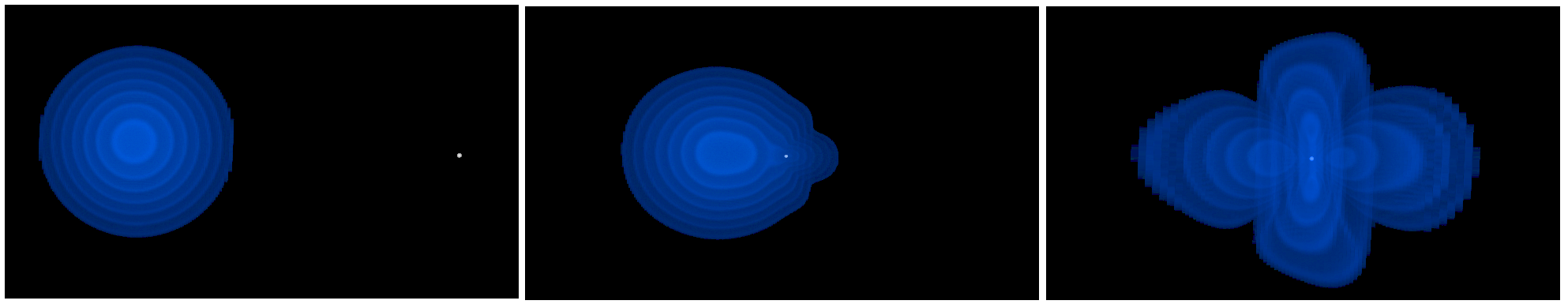}
    \caption{A volume rendering of the bosonic energy density (blue) and the BH conformal factor (grey) for the
    ${\rm BHAS}^{0.5}_{\rm 0.07}$ case for the initial, merger and final configurations. The two clouds which form are
    aligned with the x-axis along which the merger occurs.}
\label{fig:BHASs_3D}
\end{figure*}

\begin{figure*}[t]
    \centering
    \includegraphics[width=\textwidth]{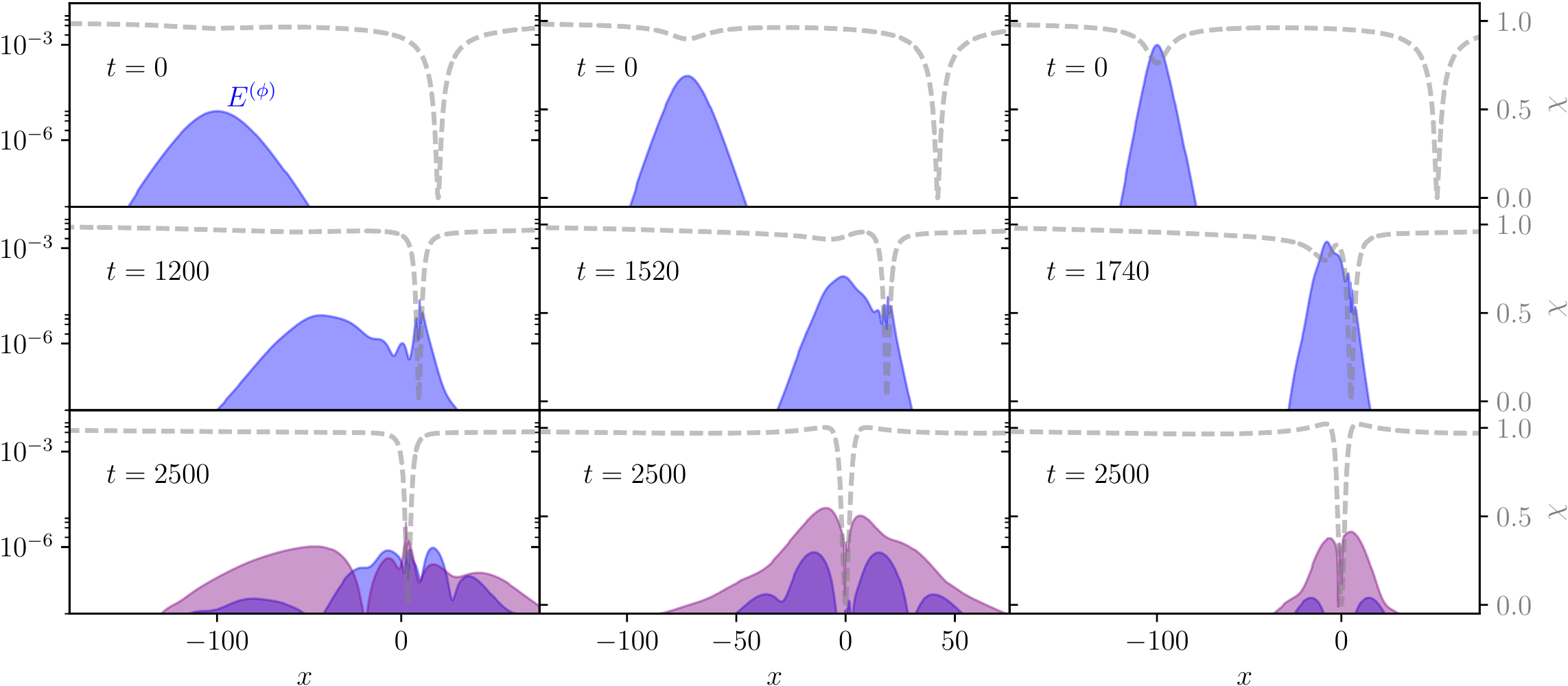}
    \caption{Evolution of the axion energy density in the non spinning (blue) and spinning (purple) cases, 
    We also show the conformal factor $\chi$ as 
    a gray dashed line, cf.~\ right axis. 
    Different rows show different instances of time 
    as labelled in each panel. We show the quantities at $t=0$, a later time corresponding approximately to the start of the merger, and the final
    timestep $t=2500 M_{\rm BH}$. 
    The different columns refer to the configurations: 
    ${\rm BHAS}^{0.0}_{\rm 0.02}$ (left), ${\rm BHAS}^{0.0}_{\rm 0.07}$ (middle), 
    ${\rm BHAS}^{0.0}_{\rm 0.20}$ (right), i.e., becoming more massive, and more compact, from left to right. 
    Only the final state is shown for the spinning case.}
\label{fig:BHASs_qualitative}
\end{figure*}

To illustrate this, we show in Fig.~\ref{fig:BHASs_qualitative}
the bosonic energy density, see Ref.~\cite{Dietrich:2018bvi}\footnote{Energy densities expressed in geometric units correspond to $1/M_{\rm BH}^2 =10^{20}~(M_\odot / M_{\rm BH})^2 ~ {\rm kg/m^3}$}, 
and the conformal factor $\chi$ for different instances of time 
for the three setups ${\rm BHAS}^{\rm 0.0}_{\rm 0.02}$ (left row), 
${\rm BHAS}^{\rm 0.0}_{\rm 0.07}$ (middle row), and ${\rm BHAS}^{\rm 0.0}_{\rm 0.20}$ (right row). 
For the final time, the result for the spinning case is also shown in purple for comparison. 
In Fig.~\ref{fig:BHASs_3D} we present a volume rendering for the ${\rm BHAS}^{\rm 0.5}_{\rm 0.07}$ case, 
to visualise the full 3D configuration of the clouds formed.

\textbf{Case I [${\rm \bf BHAS}^{\rm  \bf 0.0}_{\rm \bf 0.02}$]:}
The metric, cf.~the evolution of the conformal factor, changes only slightly once the BH and the AS merge (most of the change shown here is a gauge effect). We see that before the AS and the BH centres collide, the bosonic matter already starts to be sucked into the BH. At $t=2500 M_{\rm BH}$ the axion cloud settles to a fairly incoherent, slowly decaying configuration, with a maximum energy density of the order of $10^{-6}M_{\rm BH}^{-2}$ and a spatial extent of $\sim 100M_{\rm BH}$. A similar process occurs in the spinning case, but, most likely due to the greater dispersal of the cloud before merger due to frame dragging of the bosonic matter, the final cloud is larger than in the non spinning case, with a peak energy density of $10^{-4}M_{\rm BH}^{-2}$.

\textbf{Case II [${\rm \bf BHAS}^{\rm \bf 0.0}_{\rm \bf 0.07}$]:}
For the larger AS mass, one can see in Fig.~\ref{fig:BHASs_qualitative} that the AS is more compact with a larger central density. During the merger process, it retains more of its shape, and, since it is more compact, a greater amount is sucked into the BH once the centres meet. At $t=2500 M_{\rm BH}$ the axion cloud is more coherent than in the ${\rm BHAS}^{\rm 0.0}_{\rm 0.02}$ case. It also has a maximum energy density of the order of $10^{-6}M_{\rm BH}^{-2}$, but with a spatial extent of $\sim 50 M_{\rm BH}$, which explains the approximate order of magnitude drop in the final cloud mass shown in Fig.~\ref{fig:BHASs_clouds}. Again the spinning case results in a larger final cloud, with a peak energy density of $10^{-5}M_{\rm BH}^{-2}$.

\textbf{Case III [${\rm \bf BHAS}^{\rm \bf 0.0}_{\rm \bf 0.20}$]:}
For the largest AS mass considered, the AS is barely perturbed until the final moments of the merger and the majority of the bosonic matter falls rapidly into the BH. The remaining cloud is negligible, with a peak energy density of $10^{-7}M_{\rm BH}^{-2}$ ($10^{-6}M_{\rm BH}^{-2}$ for the spinning case), and a spatial extent of $\sim 20M_{\rm BH}$.

Overall, we see that larger clouds are formed initially for ASs with a lower mass and correspondingly lower compactness, and for higher spins. This is quantified further in the following section.

\subsection{Bosonic clouds from merger}

\begin{figure}[t]
    \centering
    \includegraphics[width=0.49\textwidth]{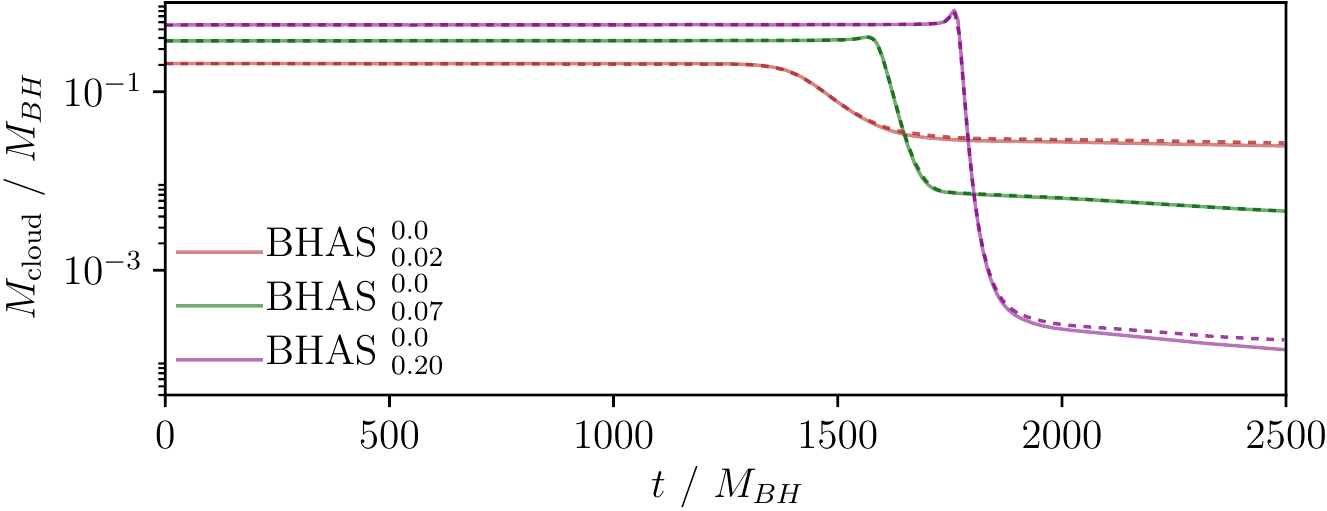}
    \caption{Evolution of the bosonic cloud mass surrounding the BH over the course of the merger. 
    We present results for resolution R3 with solid lines and corresponding results 
    for resolution R2 are shown dashed. We see that less compact ASs result in larger masses of clouds post merger, despite their initially lower mass.}
\label{fig:BHASs_clouds}
\end{figure}

\begin{figure}[t]
    \centering
    \includegraphics[width=0.49\textwidth]{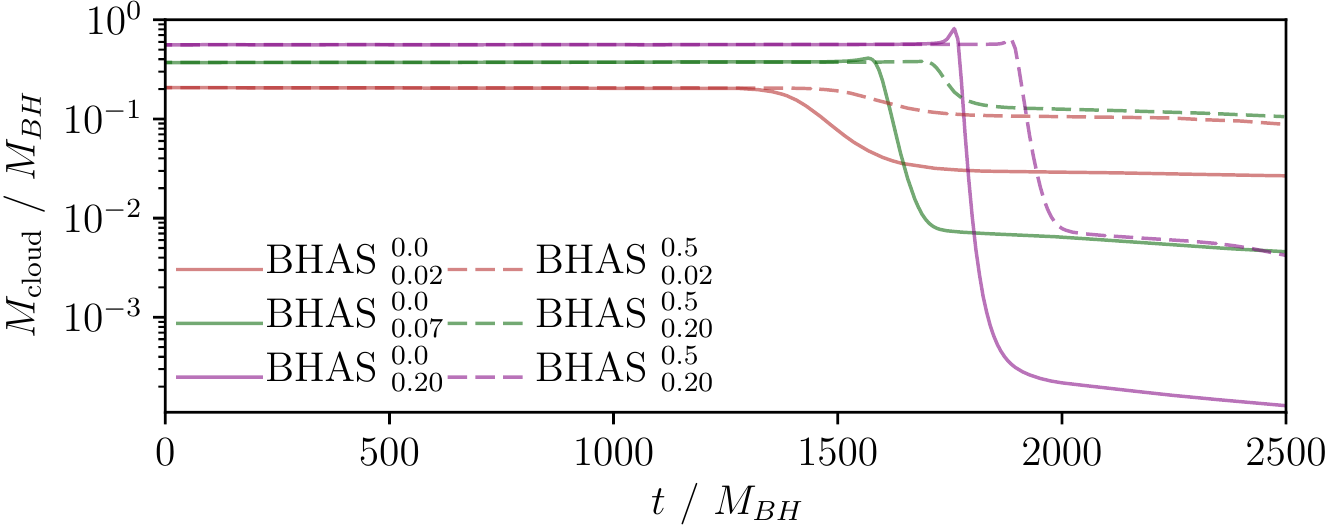}
    \caption{Evolution of the bosonic cloud mass surrounding the BH showing the effect of spin.
    We present results for resolution R3 with solid lines representing the non spinning case, $a=0$, and dashed lines showing the results
    for the corresponding case with $a=0.5$. We see that there is a clear increase 
    in the mass of the remaining cloud with spin. Note that the spin axis is perpendicular to the merger direction.
    }
\label{fig:BHASs_clouds_spin}
\end{figure}

In the case of BH-AS mergers, we are particularly interested in regions which generate large scalar field clouds, as this may lead to observable effects 
post merger, as discussed in the introduction. The cases we consider span different compactness of the AS 
(which also corresponds to adjusting the mass ratio with the BH), 
and cases with and without spin ($a=0$ and $0.5M_{\rm BH}$). 
Our results are summarised in Fig. \ref{fig:BHASs_qualitative}, Fig \ref{fig:BHASs_clouds}, and Fig. \ref{fig:BHASs_clouds_spin}.

In Fig \ref{fig:BHASs_clouds}, we show the total mass of axions over the course of the merger, obtained from integrating
\begin{equation}
M_{\rm cloud} = \int E^{(\phi)} \chi^{-3/2} dV
\end{equation}
over the numerical domain at each timestep. 
Whilst we do not explicitly excise the mass within the event horizon, 
the particular choice of gauge condition causes the matter inside the BH to fall 
into the ``puncture'' and to leave the numerical domain. Consequently, the 
majority relates to that outside the horizon. This can be clearly seen in Fig.~\ref{fig:BHASs_qualitative}. We find, which agrees to intuition, that the largest scalar clouds are generated for low compactness ASs. This appears to be mainly because the bosonic matter in the stars is more diffuse and less gravitationally bound, and thus becomes dispersed about the BH before the actual merger occurs. In the higher mass cases, the AS is more compact and retains its shape during the merger. Thus when it overlaps the BH, the majority of the mass falls immediately into the horizon.

In Fig.~\ref{fig:BHASs_clouds_spin}, 
we show the effect of the BH spin on the merger outcome. 
We find that, in general, more massive scalar clouds form from
low compactness ASs and spinning BHs. 
We observe that particular combinations appear to be more efficient at creating massive clouds, 
which might be caused by particular quasi normal mode (QNM) 
or quasi-stable bound states of the BH being excited. 

Although over 98\% of the mass is absorbed by the BH within 
1000 $M_{\rm BH}$ after the merger for most of the studied configurations, 
for some favourable cases the remaining cloud can be as 
large as 30\% of the initial AS mass, with a mass of order $10^{-1} M_{\rm BH}$, and energy densities of $\rho = 10^{-4} M_{BH}^{-2}$. This is comparable to the values obtained in superradiant build up around BHs (see \cite{Brito:2014wla}). 

As mentioned above, the results presented for BHs have a free mass parameter which means that the mass of the AS/BH system can be scaled by 
the mass of the axion. For the QCD axion $m \sim 10^{-10}$ eV, our results would correspond to a black hole of order several solar masses, 
whereas for so called ``fuzzy dark matter'' (FDM) with $m \sim 10^{-22}$ eV they would correspond to supermassive black holes.

\subsection{Gravitational wave signals}

\begin{figure}[t]
    \centering
    \includegraphics[width=0.49\textwidth]{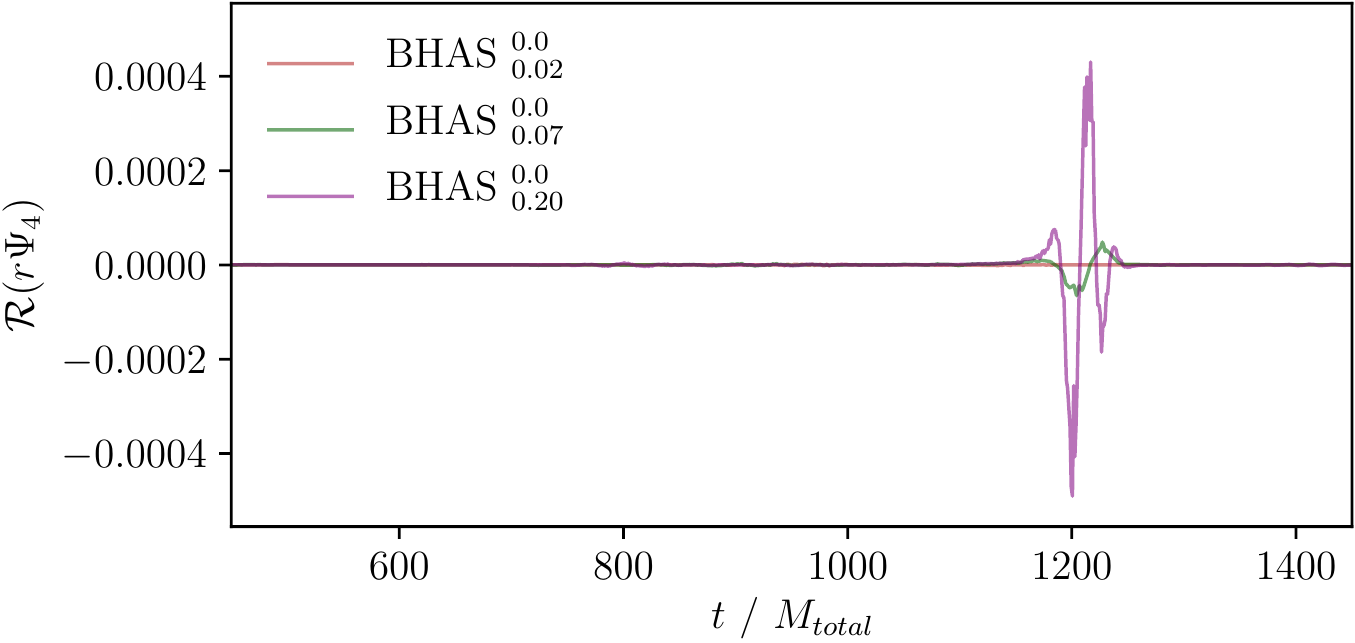}
    \caption{GW metric multipoles for the three example cases of Fig.~\ref{fig:BHASs_qualitative}, plotted on the same axis for comparison.
    The least massive case gives a barely distinguishable signal in comparison to the most compact ASs.
    }
    \label{fig:BHASs_waves}
\end{figure}

In Fig.~\ref{fig:BHASs_waves} we show the dominant (2,2)-mode of the GW signal
for the cases in Fig.~\ref{fig:BHASs_qualitative}. 
We find that for the Case I setup, e.g.\ ${\rm BHAS}_{\rm 0.02}^{0.0}$, 
the emitted GW signals emit GW energy in the (2,2)-mode of $E_{\rm GW} < 2 \times 10^{-9} $. 
For increasing AS masses the total released energy can be increased significantly. 
For the setup ${\rm BHAS}_{\rm 0.07}^{0.0}$ we obtain amplitudes about an order of magnitude larger 
than for ${\rm BHAS}_{\rm 0.02}^{0.0}$, and the emitted GW energy in the dominant mode is $E_{\rm GW}\approx 4 \times 10^{-8}$.
For the largest case considered, ${\rm BHAS}_{\rm 0.02}^{0.0}$, the amplitude is again an order of magnitude larger and the energy in the dominant mode is $E_{\rm GW}\approx 2 \times 10^{-6}$.

\section{NS-AS mergers}
\label{sec:resultsNS}

Following the study of BH-AS collisions, we will focus 
in this section on head-on mergers of NS-AS systems. 
In contrast to BHs, NSs have an intrinsic mass scale 
set by the Equation of State (EOS). 
This sets the axion mass to 
$m_a \sim 10^{-10}$ eV for the $\sim 1.38 M_\odot$ NSs which we consider, 
so that the radius of the ASs is comparable to that of the NS
(ie, since we set $\mu=1$ in geometric units).

\subsection{Configurations}

For the study of NS-AS mergers, we perform simulations 
for seven different binary configurations listed in detail in 
Table~\ref{tab:NSAS_setups}. All setups are evolved with resolutions
R1, R2, R3 as presented in Tab.~\ref{tab:BAM_grid}.
To compute the initial configurations we keep the 
bosonic and baryonic energy density fixed and solve the CTS 
equations to obtain configurations consistent with general relativity. 
Initially, we pick a NS mass in isolation of $\sim 1.38M_\odot$, 
which due to changes in the conformal factor for the binary can 
slightly increase during the 
iterative procedure to solve the Einstein's constraint equations; see~\cite{Dietrich:2018bvi} for details. 
The axion star configurations are calculated as in the BH-AS case and 
employ central values 
of the dominant zeroth mode of the scalar field of 
$\sqrt{8\pi}\phi_c^*=0.02,0.04,0.06,0.07,0.08,0.10,0.20$ 
(Table~~\ref{tab:NSAS_setups}). 
The initial distance is again adjusted such that at $t=0$ it is 
approximately $d\approx100 M_{\rm total}$. 
We employ for the simulation of the baryonic matter the 
SLy EOS~\cite{Douchin:2001sv,Read:2008iy,Dietrich:2015pxa} 
which is in agreement with most of the 
current EOS constraints directly inferred from 
GW170817~\cite{TheLIGOScientific:2017qsa,
Abbott:2018wiz,Coughlin:2018miv,Bauswein:2017vtn,
Radice:2017lry,Annala:2017llu,Most:2018hfd}.

\begin{table}[t]
  \centering    
  \caption{
    Simulated NSAS setups.
    The columns refer to: the name of the configuration, 
    the gravitational mass of the NS in isolation $M_{\rm NS}$,  
    the baryonic mass of the NS $M_{\rm NS}^*$, 
    the AS's ADM mass in isolation $M_{\rm AS}$, 
    the AS's bosonic mass in isolation $M_{\rm AS}^*$,    
    the AS's compactness in isolation $\mathcal{C} = 2M_{\rm AS} / R$     
    the central value of the axion scalar field $\phi_c(t=0)$, 
    and the initial separation $d$. } 
  \begin{tabular}{l|ccccccc}        
    \hline
    Name                    & $M_{\rm NS}$ & $M_{\rm NS}^*$ & $M_{\rm AS}$ & $M_{\rm AS}^*$ & $\mathcal{C}$ & $\sqrt{8\pi}\phi_c^*$ & $d$ \\
    \hline
    ${\rm NSAS}_{\rm 0.02}$  & 1.376 & 1.527 & 0.201 & 0.200 & 0.014 & 0.02 & 155.1 \\  %0.2074
    ${\rm NSAS}_{\rm 0.04}$  & 1.380 & 1.532 & 0.279 & 0.277 & 0.028 & 0.04 & 162.9 \\  % 0.2891
    ${\rm NSAS}_{\rm 0.06}$  & 1.383 & 1.536 & 0.334 & 0.330 & 0.042 & 0.06 & 168.4 \\  %0.3478
    ${\rm NSAS}_{\rm 0.07}$  & 1.384 & 1.537 & 0.357 & 0.352 & 0.048 & 0.07 & 172.8 \\  %0.3723
    ${\rm NSAS}_{\rm 0.08}$  & 1.385 & 1.538 & 0.378 & 0.371 & 0.055 & 0.08 & 172.8 \\  %0.3947
    ${\rm NSAS}_{\rm 0.10}$  & 1.387 & 1.540 & 0.413 & 0.404 & 0.068 & 0.10 & 176.3 \\  %0.4337   
    ${\rm NSAS}_{\rm 0.20}$  & 1.391 & 1.546 & 0.525 & 0.502 & 0.130 & 0.20 & 187.4\\   %0.5637     
    \hline
  \end{tabular}
  \label{tab:NSAS_setups}
\end{table}

\subsection{Qualitative merger dynamics}

\begin{figure*}[t]
    \centering
    \includegraphics[width=\textwidth]{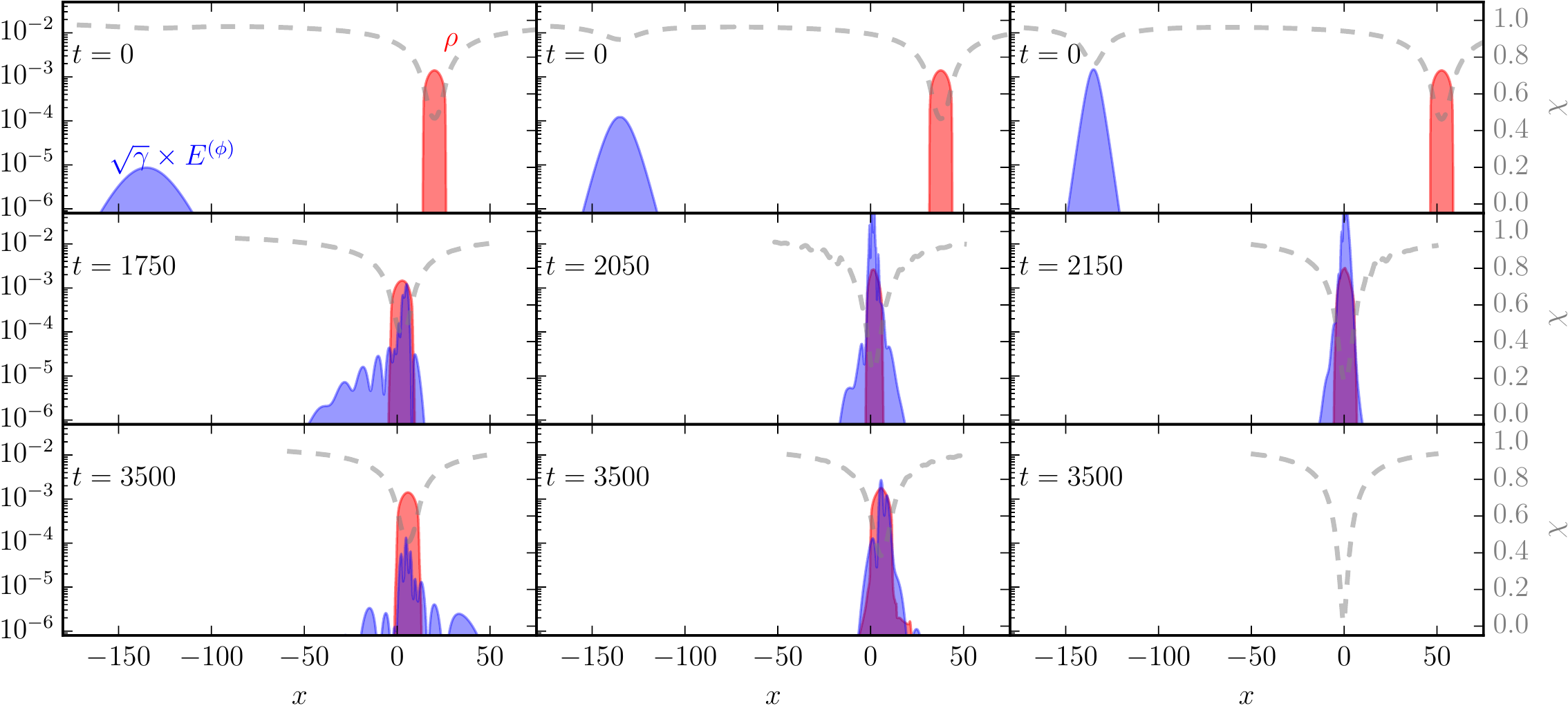}
    \caption{Evolution of the bosonic energy density (blue) 
    incorporating the determinant of the 3-metric to allow the computation 
    of the bosonic cloud (Fig.~\ref{fig:NSASs_disks}) and 
    the baryonic density (red). We also present the conformal factor $\chi$ as 
    a grey dashed line, cf.~\ right axis. 
    We show the conformal factor for $t=0$ at level $l=3$
    and for later times at refinement level $l=4$. 
    The different columns refer to the configurations: 
    ${\rm NSAS}_{\rm 0.02}$ (left), ${\rm NSAS}_{\rm 0.07}$ (middle), 
    ${\rm NSAS}_{\rm 0.20}$ (right). Different columns show different instances of time 
    as labeled in each panel. }
\label{fig:NSASs_qualitative}
\end{figure*}

We can classify the merger dynamics of NS-AS systems in 
three qualitatively different categories:
\begin{enumerate}[(I)]
\item In cases for which the AS has a small mass (and correspondingly low compactness), 
the NS is only weakly perturbed 
during the collision. Thus, only a small amount of matter is ejected and GW luminosity is small. 
The NS stabilises and becomes surrounded by a bosonic cloud which appears to be long lived. 
\item Increasing the AS mass, we obtain a merger remnant which is strongly excited, thus
emitting GWs and ejecting large amounts of baryonic matter. 
\item In cases where the axion star is even more massive, 
the final remnant is a black hole. For these cases the GW luminosity is larger, 
but baryonic and bosonic ejecta are suppressed.
\end{enumerate}

\begin{figure}[t]
    \centering
    \includegraphics[width=0.49\textwidth]{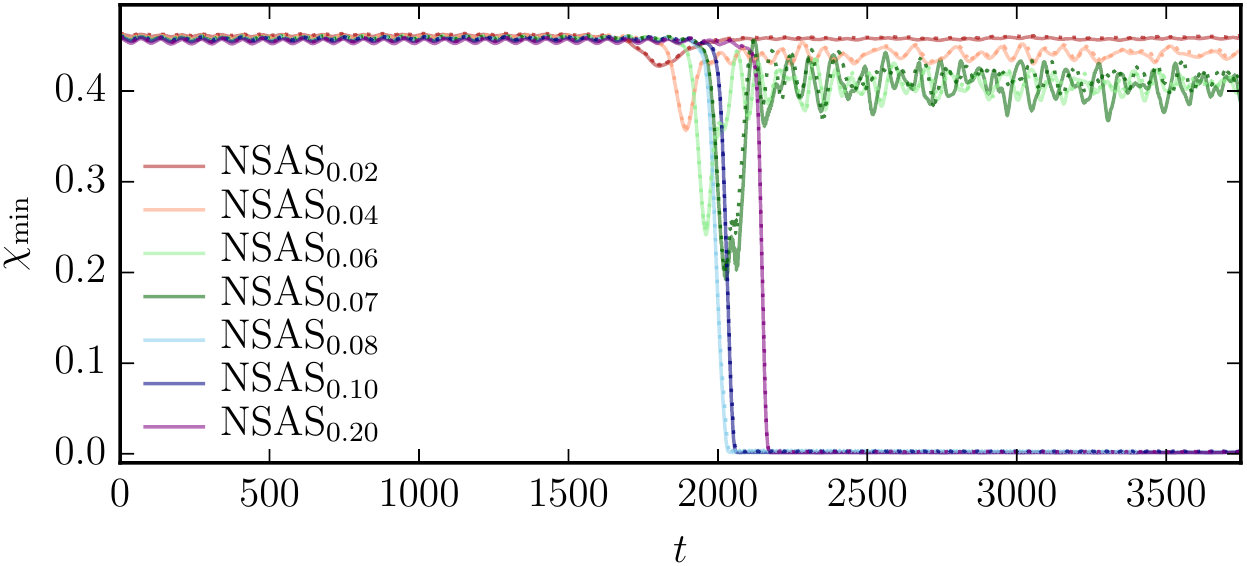}
    \caption{Evolution of the minimum value of the conformal factor $\chi$. 
    We present results for resolution R3 with solid lines and corresponding results 
    for resolution R2 are shown dashed. Overall we find good agreement between 
    different resolutions showing the robustness of the numerical methods.}
\label{fig:NSASs_chi}
\end{figure}

In the following we discuss in more detail
these possible merger outcomes. For this purpose we show in Fig.~\ref{fig:NSASs_qualitative}
the bosonic and baryonic energy density, see Ref.~\cite{Dietrich:2018bvi}, 
and the conformal factor $\chi$ for different times 
for the three setups ${\rm NSAS}_{\rm 0.02}$ (left row), 
${\rm NSAS}_{\rm 0.07}$ (middle row), and ${\rm NSAS}_{\rm 0.20}$ (right row). 
Additionally, we also present the time evolution of the conformal factor in 
Fig.~\ref{fig:NSASs_chi}. 
This can be compared to the same cases for BHAS mergers, in Fig.~\ref{fig:BHASs_qualitative}.

\textbf{Case I [${\rm \bf NSAS}_{\rm \bf 0.02}$]:}
From the upper left panel we can immediately see qualitatively differences 
between bosonic and baryonic stars. 
While ASs do not have a sharp surface and are characterized by an 
exponentially decaying scalar field, the density at the NSs surface 
is only $C^0$ continuous. 
Considering ${\rm NSAS}_{\rm 0.02}$ we find that initially 
the central energy density of the AS 
is about $\sim 2$ orders of magnitude below the central density of the NS. 
However, the peak energy density significantly increases during the merger process 
becoming compatible to the NS's density. 
Nevertheless, the overall change in the metric, 
cf.~the evolution of the conformal factor, 
changes only slightly once the NS and the AS merge. 
In fact at around $t\approx 2000$ the minimum of the conformal 
factor returns to the initial value before the collision; 
see Fig.~\ref{fig:NSASs_chi}. 
Finally at $t=3500$ the NS settles to a setup for which 
it is surrounded by a bosonic cloud, 
which again has maximum energy densities 
of the order of $10^{-4}$. 

\textbf{Case II [${\rm \bf NSAS}_{\rm \bf 0.07}$]:}
For cases where the AS's mass is larger than for Case I setups, 
we find that at merger the central bosonic energy density 
can be about $2$ order of magnitude larger. 
The central value of the conformal factor 
decreases from $\sim 0.45$ to $0.2$, 
but no BH forms during the merger process. 
At this stage the NS's central density increases and 
the radius of the NS shrinks, see middle panel 
of Fig.~\ref{fig:NSASs_qualitative}. 
We expect that the transition between BH formation and no 
BH formation is characterized by a type-I critical
phenomenon as in the NSNS merger case~\cite{Kellermann:2010rt}. 
We postphone a careful investigation employing a larger number
of configurations for future work.
Even at $t=3500$ the remnant has not yet settled to a stable configuration 
and undergoes continuous oscillations (see green line in Fig.~\ref{fig:NSASs_chi}). 
Compared to the Case I setup, we find larger differences 
between resolutions R2 and R3, however, the qualitative shape agrees well, 
emphasising the robustness of the results. 

\textbf{Case III [${\rm \bf NSAS}_{\rm \bf 0.20}$]:}
As the AS mass is further increased we find that 
(i) ASs are more compact
with energy densities compatible with the NS density. 
In contrast to Case II setups, the pressure of the bosonic and baryonic matter 
is not able to counteract the gravitational collapse, which is clearly 
visible by the conformal factor becoming zero. 
At $t=3500$ we find that for such cases there is no noticeable 
baryonic disk surrounding the final remnant and also the bosonic cloud is 
relatively small.

\subsection{Bosonic cloud and baryonic outflow}

\begin{figure}[t]
    \centering
    \includegraphics[width=0.49\textwidth]{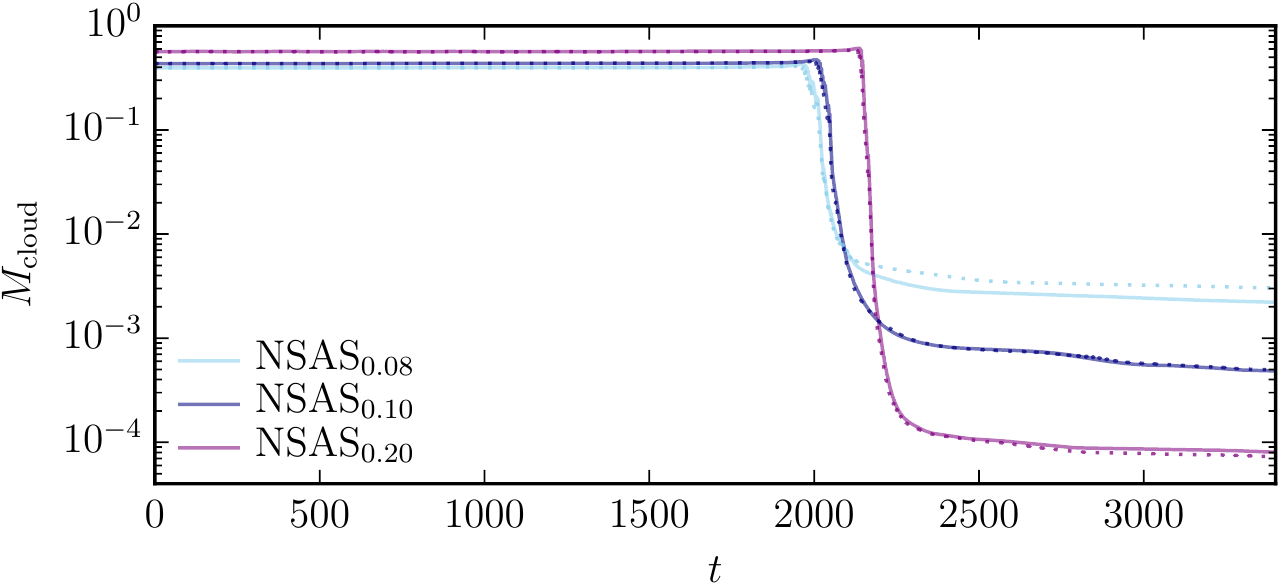}
    \caption{Evolution of the bosonic cloud mass surrounding the final BH remnant. 
    We present results for resolution R3 with solid lines and corresponding results 
    for resolution R2 are shown dashed. 
    We evaluate the cloud mass on the refinement level $l=1$. }
\label{fig:NSASs_disks}
\end{figure}

As for the BHAS mergers, we are interested in the potential formation of a 
BH surrounded by a cloud of bosonic particles. 
We present in Fig.~\ref{fig:NSASs_disks} the cloud mass for the cases which 
undergo BH formation. In agreement with our previous investigations 
we find that the cloud mass is larger for the case of 
less compact ASs. 
In particular for the ${\rm NSAS}_{\rm 0.08}$ configuration 
the bosonic cloud reaches up to $\sim 2\times 10^{-3} M_\odot$. 
We assume that close to the threshold of BH formation 
there will be a maximum achievable BH axionic cloud mass.
We expect that as presented for the study of BH-AS systems 
an additional intrinsic spin of the NS or also the consideration of 
systems with orbital angular momentum will increase the cloud 
and disk mass. 

Considering the amount of baryonic mass surrounding the final remnant, we find that even shortly after BH formation the baryonic disk mass drops below 
$<10^{-8} M_\odot$. Such small disk masses would generally 
not give rise to short gamma-ray-bursts (GRBs).
We assume that the reason for this small disk mass is 
the missing angular momentum support due to the fact that we restrict 
our analysis to head-on collisions. 
Further studies investigating orbiting 
NS-AS simulations will be needed. 
 
\begin{figure}[t]
    \centering
    \includegraphics[width=0.49\textwidth]{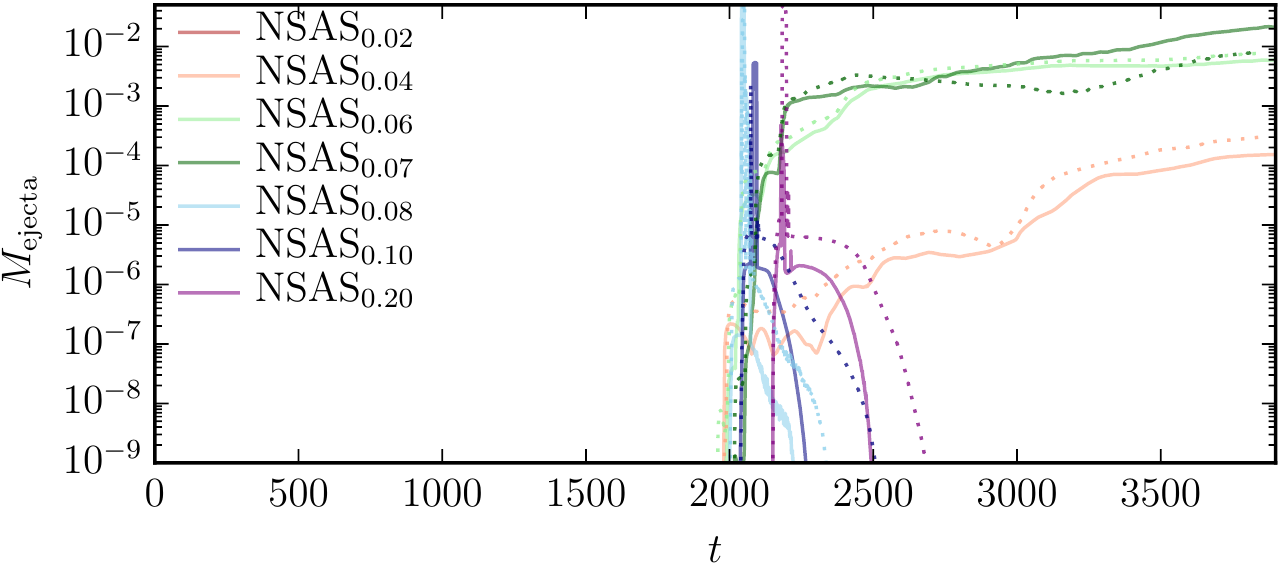}
    \caption{Evolution of the baryonic matter ejected during and after the merger process of the NSAS systems. 
    Results for resolution R3 are shown solid, corresponding results 
    for resolution R2 are dashed. 
    We evaluate the ejecta mass on the refinement level $l=1$.}
\label{fig:NSASs_Du}
\end{figure}

In addition to the baryonic disk acting as the central engine 
for a possible sGRB, another electromagnetic counterpart 
might be triggered by the neutron-rich outflow of the baryonic matter, 
namely a kilonova~\cite{LaSc1974,LiPa1998,MeMa2010,RoKa2011,KaMe2017}. 
The kilonovae properties depend on the ejecta mass, geometry, 
and composition. Generally, more massive ejecta are able to trigger 
brighter kilonovae. Therefore an estimate of the ejected baryonic mass 
is of crucial importance. 
We show the ejecta mass estimates for all studied configurations 
in Fig.~\ref{fig:NSASs_Du}.
For systems close to the threshold of BH formation 
the amount of ejected matter easily exceeds $10^{-2}M_\odot$. 
Consequently this material should be able to create electromagnetic 
signatures which we will describe in detail in~\cite{Dietrich:inprep}. 

\subsection{BH formation threshold}

\begin{figure}[t]
    \centering
    \includegraphics[width=0.49\textwidth]{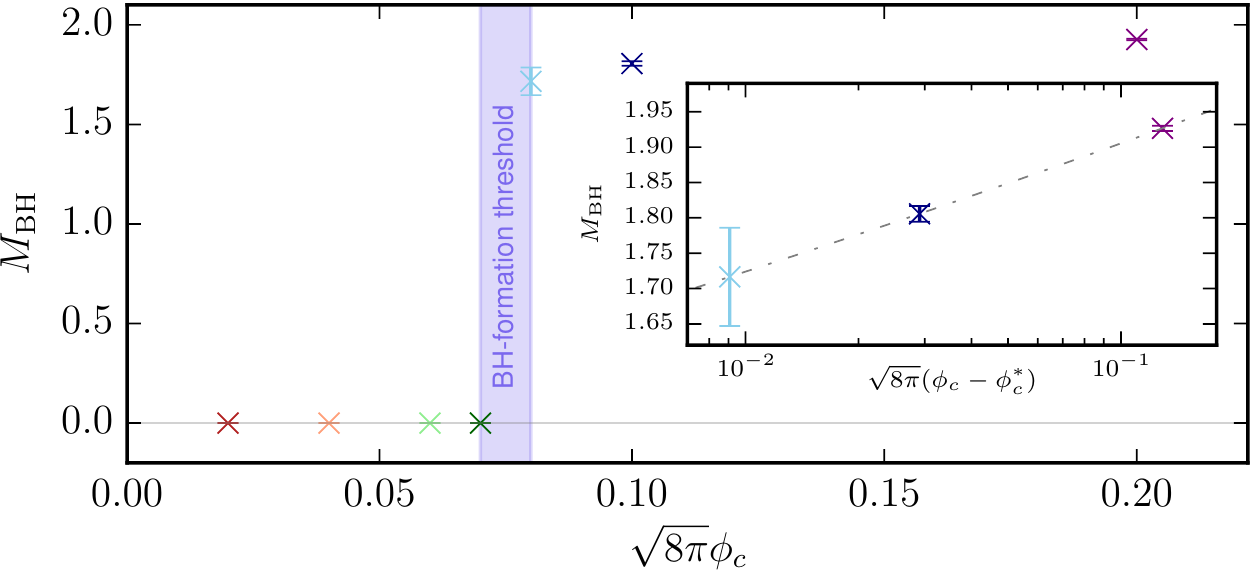}
    \caption{BH mass as a function of the initial central value of the AS.
    For the inset we show the black hole mass for supercritical configurations 
    assuming a critical BH formation threshold at $\sqrt{8\pi}\phi_c^*=0.0709$.
    Error bars refer to the difference between the resolutions R3 and R2. }
    \label{fig:NSASs_horizons}
\end{figure}

With reference to the threshold of BH formation 
for the configurations in this section, we present 
the BH mass as a function of the initial central value of the AS in Fig.~\ref{fig:NSASs_horizons}.
As can be concluded from Fig.~\ref{fig:NSASs_chi} 
the critical value for BH formation lies in 
$\sqrt{8\pi}\phi_c^* \in [0.07,0.08]$. 
Motivated by studies of the critical collapse, 
see~\cite{Gundlach:2007gc} and references therein, we assume that 
the BH mass is proportional to
\begin{equation}
    M_{\rm BH} \propto A \left[ (\phi_c - \phi_c^*) \right]^\delta. 
\end{equation}
We fit the supercritical configurations to this function and  
obtain an aproximate threshold of BH formation of $\sqrt{8\pi}\phi_c^*=0.071$, as shown in Fig.~\ref{fig:NSASs_horizons}.
However, we would require many more simulations closer to the critical point to obtain accurate values. 
For type II collapses, as $\phi_c \rightarrow \phi_c^*$
the BH mass should approach zero, and thus
there is the possibility that at the threshold of BH formation 
all the baryonic and bosonic matter gets ejected from the system. 
However, we would expect that the collapse is type I, as is the case in NS-NS mergers~\cite{Kellermann:2010rt}.

\subsection{Gravitational wave signal}

\begin{figure}[t]
    \centering
    \includegraphics[width=0.49\textwidth]{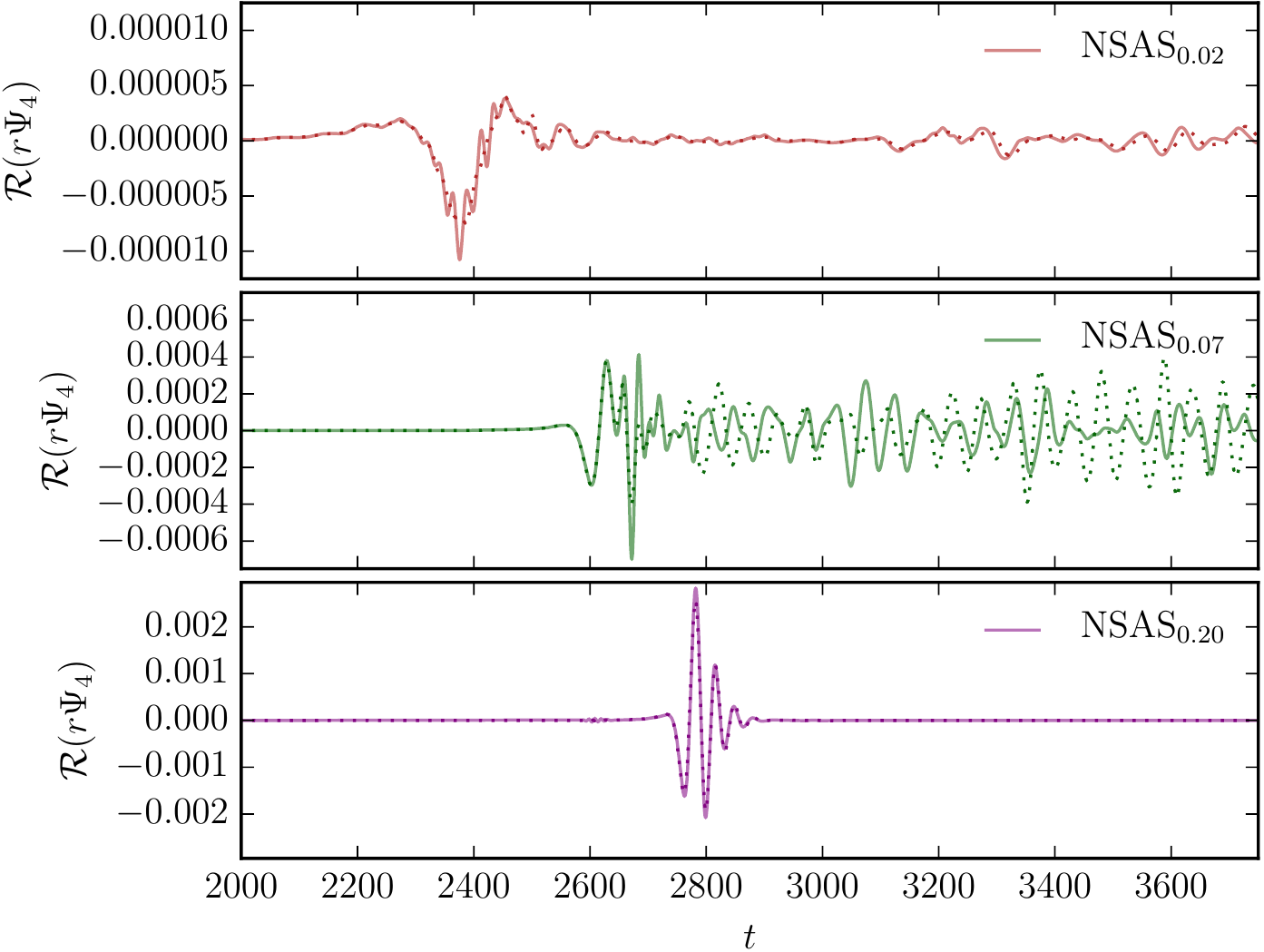}
    \caption{GW metric multipoles for the three example cases of Fig.~\ref{fig:NSASs_qualitative}. 
    Results for resolution R3 are shown solid, corresponding results 
    for resolution R2 are dashed. }
    \label{fig:NSASs_waves}
\end{figure}

We close our discussion about NS-AS mergers by considering the emitted 
GW signal. The dominant (2,2)-mode of the GW signal is shown in Fig.~\ref{fig:NSASs_waves}
for the cases in Fig.~\ref{fig:NSASs_qualitative}. 
We find that for the Case I setup, e.g.\ ${\rm NSAS}_{\rm 0.02}$, 
the emitted GW signals have amplitudes of about $10^{-5}$. 
Over the entire simulation the emitted GW energy is $E_{\rm GW} \lesssim 1\times10^{-5}$. 
For increasing AS masses the total released energy can be increased by orders of magnitudes. 
For setup ${\rm NSAS}_{\rm 0.07}$ we obtain amplitudes more than an order 
of magnitude larger than for ${\rm NSAS}_{\rm 0.02}$ and the emitted 
GW energy is about $E_{\rm GW}\approx 2\times 10^{-4}$. Furthermore, 
the remnant is still highly dynamical and the GW luminosity has not decreased 
noticeably by the end of the simulation. 
Due to the highly dynamical postmerger regime we find that 
overall agreement between different resolutions
is worse than for ${\rm NSAS}_{\rm 0.02}$, cf.~dashed and solid lines. 
However, different resolutions (including resolution R1) give similar 
results and lead to the same estimate of the emitted GW energy. 
Finally, for cases which form BHs during the evolution
the amount of GW energy can grow up to 
$5 \times 10^{-4}$. 
As an example we show the case for ${\rm NSAS}_{\rm 0.20}$. 

\section{\label{sec:conclusions}Discussion}

In this article we have presented what is, to the best of our knowledge,
the first study of axion star collisions with black holes 
and neutron stars using full 3+1D numerical relativity simulations. 
Such a study seems timely in the gravitational wave 
astronomy era, in which multiple detections of compact binaries 
are expected in coming years~\cite{Abbott:2016nhf}.

With respect to our black hole-axion star merger simulations, we have investigated the impact 
of the axion star's compactness, and the BH spin, on the mass of the remnant bosonic cloud surrounding the 
black hole. 
Although in most of the considered cases $\sim 98\%$ of the axion star's mass 
is absorbed by the black hole shortly after the merger, in favourable cases the 
remaining cloud can be as large as $30\%$ of the initial axion star mass, 
with a bosonic cloud of mass of $\mathcal{O}(10^{-1})M_{\rm BH}$ and peak energy density of $10^{-4}$, 
comparable to that obtained in a superradiant build up.
We find that the largest scalar clouds are generated for 
low compactness ASs and spinning black holes. 
We note that there appear to be particular combinations which are overall more  
efficient at producing large axion clouds. We speculate that this might be caused by 
the excitation of particular quasi-bound states of the black hole.

The presented results are important
since they show (i) that axion star-black holes mergers can provide a dynamical mechanism for 
the formation of scalar hair around black holes and
(ii) that faster spinning (but not yet extremal) black holes allow for relatively large cloud masses. 
The spinning case is especially interesting as it may provide the seed for a superradiant build up,
which could lead to additional observable gravitational wave signatures post merger. However, superradiance 
requires extremal spins and an appropriate matching of the axion and BH masses, whereas the effects we observe here are
in principle more general.
It would be worth extending this study to a larger range of mass ratios, spins and spin orientations, to confirm the approximate trends 
observed in this paper and identify 
whether the proposal that particular mass ratios and spin combinations are favourable for forming clouds is consistent with a wider set of results. 
It would also be interesting to consider the effect of larger self interactions of the axion field, 
other values of $M\mu$ and, in the longer term, 
interactions with baryonic matter in an accretion disc.\\

For our study of neutron star-axion star collisions, we restricted our investigations to the merger of  
axion stars of various compactnesses with a ``typical'' neutron star having 
a gravitational mass of $\sim 1.38M_\odot$ and the SLy equation of state. 
We found that for the setups studied, there exists a critical mass threshold
for the axion star required to form a BH during the collision. 
In the considered cases, the black hole formation is triggered by the axion star being 
perturbed within the potential well of the neutron star.
Its collapse leads to a black hole within the 
neutron star, rather than collapse of the neutron star itself.

For sub-threshold axion star masses the merger remnant is a 
perturbed neutron star enveloped in an axion cloud. 
For super-threshold axion star masses the final remnant is a black hole 
with a scalar cloud surrounding it. 
We suggest that the black hole formation threshold may correspond to a type I critical 
phase transition, as in binary neutron star mergers, and therefore universality and scaling relations could exist near to the critical point. 
We present a first (although very approximate) estimate of
the critical threshold parameter $\phi_c^*$, but
further simulations are required for a more stringent constraints on the critical parameters.

Interestingly, we found that in the marginally sub critical cases, 
a large amount of baryonic mass was released from the merger remnant due to 
the formation of shocks in the NS. These ejecta can give rise to a kilonova-like counterpart, such as ATF201gfo, e.g.~\cite{Monitor:2017mdv,Coulter:2017wya,Cowperthwaite:2017dyu,Smartt:2017fuw,Kasliwal:2017ngb,Kasen:2017sxr}. 
The potential new type of transient produced by such a near-critical neutron star-axion star collision is discussed in 
more detail in~\cite{Dietrich:inprep}. 
In cases where a black hole forms after the merger, 
the ejection of matter as well as the formation of a 
baryonic accretion disk or bosonic cloud is suppressed. 
However, in the most extreme case the final black hole remnant can be embedded in a 
bosonic cloud of mass $\mathcal{O}(10^{-3})M_\odot$.
 
In future, we plan to perform further numerical simulations in which we add a direct interaction between 
the axions and the neutron star fluid. Such couplings, which are necessary to correctly model the QCD axion, 
could also give rise to observable effects 
in NS-NS collisions that occur in a background of axions.
Such an approach would allow us to further constrain the properties of axionic dark matter
using observations of the merger of binary neutron stars within dark matter halos. 

\section{\label{sec:acknowledge}Acknowledgements}

KC thanks her GRChombo collaborators (\url{www.grchombo.org/collaborators}), 
and masters student Sladja Radnovic for her initial work on the simulations.
TD acknowledges support by the European Unions Horizon 2020 research and 
innovation program under grant agreement No 749145, BNSmergers.
Computations have been performed on the supercomputer SuperMUC at the LRZ
(Munich) under the project number pr48pu, 
the compute cluster Minerva of the Max-Planck Institute for Gravitational Physics,
BSC Marenostrum IV via PRACE grant Tier0 PPFPWG, La Palma Astrophysics Centre 
via BSC/RES grants AECT-2017-2-0011 and AECT-2017-3-0009 and on SurfSara 
Cartesius under Tier-1 PRACE grant DECI14 14DECI0017, and the GWDG cluster in Goettingen.

\bibliography{paper20180814.bbl}

\end{document}